\documentclass[11pt]{article}
\usepackage{amsfonts}
\usepackage{amssymb}
\usepackage{amsmath}
\usepackage{geometry}
\usepackage{graphicx}
\usepackage{color}

\newtheorem{theorem}{Theorem}

\def\singlespacing{\baselineskip=13pt}	
\begin{document}

\title{Elastodynamics of radially inhomogeneous spherically anisotropic elastic materials   in the Stroh formalism }  

\author{A.N. Norris$^{a}$ and  A.L. Shuvalov$^{b}$  \\ ~ \\ 
$^{a}$ Mechanical and Aerospace Engineering, Rutgers University, \\
Piscataway, NJ 08854-8058, USA \\
$^{b} $Universit\'{e} de Bordeaux, Institut de M\'{e}canique et
d'Ing\'{e}nierie de Bordeaux, \\
UMR 5295, Talence 33405, France
} 


\maketitle

\begin{abstract}

A method is presented for solving elastodynamic problems in radially inhomogeneous elastic materials with spherical anisotropy, i.e.\ materials such that $c_{ijkl}= c_{ijkl}(r)$ in a spherical coordinate system $\{ r,\theta ,\phi \}$.  
The time harmonic displacement field $\mathbf{u}(r,\theta ,\phi )$ is expanded in  a separation of variables form with  dependence on $\theta ,\phi$ described by vector spherical harmonics with 
$r$-dependent amplitudes. 
It is proved that such separation of variables solution is generally possible only if the spherical anisotropy is restricted to transverse isotropy with the principal  axis in the radial direction, in which case the amplitudes are determined by a  first-order ordinary differential system.   
Restricted forms of the displacement field, such as $\mathbf{u}(r,\theta )$, admit this type of separation of variables solutions for certain lower material symmetries.  These results extend the Stroh formalism of elastodynamics in rectangular and cylindrical  systems to spherical coordinates. 

\end{abstract}

\section{Introduction}  \label{sec1}

The Stroh formalism \cite{Strohbook}, which recasts equations of
time-harmonic elastodynamics in the form of a first-order ordinary
differential system (ODS), is a powerful technique for dealing with elastic
materials  inhomogeneous in one coordinate.  The method was
originally established for rectangular coordinates, e.g. \cite%
{Ting96,Wu98,Shuvalov00a,Shuvalov04a}, and has been developed for
applications in cylindrical coordinate systems \cite{Shuvalov03,Norris10}.
One complicating factor for cylindrical, as compared with rectangular,
anisotropy is that the radial and azimuthal basis vectors $\mathbf{e}_{r}$
and $\mathbf{e}_{\theta }$ depend upon the angular coordinate $\theta $.
This however does not hamper separation of variables and allows for a
Stroh-like ODS provided that the material coefficients depend either on the
radial or axial coordinate $r$ or $z$.  
The situation   is quite different for spherical  anisotropy.  
The lowest anisotropy that supports general  displacement fields 
which can be described by a separation of variables appears to be transverse isotropy (TI) with the axis of symmetry in the radial direction, as shown by \cite{Hu54} for statics and \cite{Shulga88} for dynamics ((the term "spherically isotropic" used by \cite{Hu54} and others is equivalent to TI).    A state space system was developed for radially inhomogeneous  TI by \cite{Shulga88},  who also identified two distinct types of  wave motion solutions:  an uncoupled pure shear  motion and  a coupled  radial-angular solution pair.    The state vector approach has been applied    to vibrations of thick-walled TI shells \cite{Chen01,Hash09} and further developed for  piezoelectric shells \cite{Scandrett02}.   Lower symmetry can support specific types of kinematically restricted deformation.  The most general form of spherical anisotropy which admits static solutions of the form $\mathbf{u} =u(r)\mathbf{e}_{r}$ in spherical coordinates $(r,\theta ,\phi )$ is
described in \cite{Ting98}.

The purpose of this paper is to present a method for solving elastodynamic
problems in radially inhomogeneous elastic materials with spherical
anisotropy, i.e.\ materials such that $c_{ijkl}= c_{ijkl}(r)$ in
a spherical coordinate system \cite{Lekhnitskii}. No \textit{a priori}
restrictions are made on the form of the displacement field. The main departure from previous works is the use of vector spherical harmonic functions  as the set of vector  basis functions. We  show that the most general type of spherical symmetry, for which the
basis of vector spherical harmonics always yields a separable solution, is
transverse isotropy about $\mathbf{e}_{r}$ (certain lower symmetries admit
such solutions as well but only under appropriate kinematic restrictions).
Such anisotropy restriction may actually be not so severe, since any
spherical anisotropic material with properties independent of the
polar-axis orientation must possess  transverse isotropy anyway (see
below).   
The spherical TI problem reduces to an ODS in the radial coordinate $r$ with a system matrix possessing  Hermiticity properties that  guarantee physical attributes such as energy conservation. 
The key feature of the present analysis is the set of basis functions, vector spherical harmonics, which  allow for the first time application of the full Stroh
formalism to spherical elasticity for arbitrary displacement fields. 

The paper is organized as follows. The concept of spherical anisotropy is
revisited in \S \ref{sec1.1}.  Vector
spherical harmonic basis functions are introduced in \S \ref{sec2}. The main
result, which is the Stroh-like ODS for the state vector comprising the $r$%
-dependent components of displacement and radial traction in the basis of
vector spherical harmonics, is described in \S \ref{sec3} and 
proved in detail in \S \ref{sec4}. Explicit solutions and their properties
are discussed in \S \ref{sec5}. Conclusions  and further prospects are presented in \S \ref%
{sec6}.

\section{Elastic anisotropy in cylindrical and spherical coordinates}

\label{sec1.1}

The concept of cylindrical and spherical elastic anisotropy was introduced by Saint-Venant and subsequently developed by Lekhnitskii \cite%
{Lekhnitskii}. It is motivated by the existence of special materials
possessing either a physically distinguished direction aligned with the axis 
$Z$ of the cylindrical coordinate system or a point which can be identified
as the origin $O$ of the spherical system. It is physically relevant to
consider such materials in terms of tensor fields in an orthogonal  curvilinear
coordinate system.  
A tensor ${\pmb\Phi }$ of order $p\in {\mathbb N}$ is defined  
at every point $\mathbf{r}$ of a material body $\mathcal{P}$
by the array  $\left\{ \Phi _{i_{1}..i_{p}}\right\} $  
$(i_{j}=1,2,3;\ j=1,...,p) $  of its  components 
in some frame of orthogonal basis vectors $\left( \mathbf{e}_{1},\mathbf{e}_{2},\mathbf{e}_{3}\right) $ associated with 
$\mathbf{r}\in {\mathbb R}^3$.  The components change in the usual manner under a change of 
basis (see e.g. \cite[\S 2.6]{Nair}). 
Recall that the frame $\left\{ \mathbf{e}\right\}
=\left( \mathbf{e}_{x},\mathbf{e}_{y},\mathbf{e}_{z}\right) $ of a
rectangular (orthogonal rectilinear) coordinate system in ${\mathbb R}^3$ is
independent of $\mathbf{r}$ and of the origin point $O\mathbf{,}$ whereas an
orthogonal curvilinear coordinate system implies a varying frame, e.g., the
frame $\left\{ \mathbf{e}\left( \theta \right) \right\} =\left( \mathbf{e}%
_{r},\mathbf{e}_{\theta },\mathbf{e}_{z}\right) $ with a fixed longitudinal
axis $Z$ $\parallel $ $\mathbf{e}_{z}$ for the cylindrical system, and the
frame $\left\{ \mathbf{e}\left( \theta ,\phi \right) \right\} =\left( 
\mathbf{e}_{r},\mathbf{e}_{\theta },\mathbf{e}_{\phi }\right) $ with a fixed
origin $O$ and polar axis $Z$ for the spherical system. The essential
difference between the rectilinear and curvilinear arrays of components  
    is related to the meaning of a uniform tensor, i.e., one 
independent of $\mathbf{r}$. If the array of components $\left\{ \Phi _{i_{1}..i_{p}}\right\} $ in some 
rectangular frame is uniform: 
$\Phi _{i_{1}..i_{p}}=$const.\ for all $i_{j},j$ and $\mathbf{r}\in \mathcal{P}$,
then the tensor  ${\pmb\Phi }$  is also uniform,
i.e.\ is the same for all $\mathbf{r}$, and vice versa. By contrast, a
uniform array of components  $\left\{ \Phi _{i_{1}..i_{p}}\right\} $ referred to a
curvilinear frame $\left\{ \mathbf{e}\right\} $ describes a  tensor  
${\pmb \Phi }$ which generally depends upon $\mathbf{r}$ since    $\left\{ 
\mathbf{e}\right\} $ does. 
For example, ${\pmb\Phi }={\pmb\Phi }\left( \theta \right) $  for a
uniform  cylindrical array of components and 
${\pmb\Phi }={\pmb\Phi } \left( \theta ,\phi \right) $   for uniform spherical components, 
 unless these tensors are isotropic (components invariant to change of basis under $SO\left( 3\right) $) or if the
cylindrical one is transversely isotropic (invariant under $SO\left( 2\right) $
about $\mathbf{e}_{z}$). As a matter of definition, a   tensor ${\pmb\Phi }$ 
associated with   cylindrical or spherical   components  
which is not
isotropic in the above sense, is called \textit{cylindrically or spherically
anisotropic}, respectively. 

The components of a tensor  may certainly be non-uniform so that $\left\{ \Phi
_{i_{1}..i_{p}}\right\} =\left\{ \Phi _{i_{1}..i_{p}}\left( \mathbf{r}%
\right) \right\} $ (where $\mathbf{r}$ does not have to be related to the
same coordinate system as the frame $\left\{ \mathbf{e}\right\} $). Many
applications deal with a specific type of inhomogeneous body $\mathcal{P}$
in which the curvilinear components $\Phi _{i_{1}..i_{p}}( \mathbf{r} ) $ of a
non-uniform  tensor vary with position but maintain the same
anisotropy (symmetry class) at every $\mathbf{r}\in \mathcal{P}$. This is
the general framework considered in this paper. A widely encountered example
is the case of a cylindrical or spherical radially inhomogeneous elastic
tensor with $c_{ijkl}=c_{ijkl}\left( r\right) $, where $r$ may be measured
from any point of the fixed axis $Z$ of the cylindrical system (since $%
c_{ijkl}$ do not depend on $z$) or from the fixed origin $O$ of the
spherical system.

Another particular aspect of cylindrical and spherical  tensor  components   $\left\{ \Phi
_{i_{1}..i_{p}}\right\} $ stems from the fact that the {\em orientation} of frame
vectors is undefined at certain points, namely, (i) of the pair $\left( 
\mathbf{e}_{r},\mathbf{e}_{\theta }\right) $ of the cylindrical frame at the
points $r=0$ lying on the longitudinal axis $Z$, (ii) of the pair $\left( 
\mathbf{e}_{\theta },\mathbf{e}_{\phi }\right) $ of the spherical frame at
the points $\theta =0,\pi $ of the polar axis $Z$, and (iii) of the whole
spherical frame $\left( \mathbf{e}_{r},\mathbf{e}_{\theta },\mathbf{e}_{\phi
}\right) $ at the point $r=0$ of the origin $O$. Now suppose that a given
body $\mathcal{P}$ contains either the axis $Z$ or the point $O$ and has a
tensor ${\pmb\Phi }$   described by the array of components
$\left\{ \Phi _{i_{1}..i_{p}} (\mathbf{r} ) \right\} $ which is single-valued at every $%
\mathbf{r}\in \mathcal{P}$ including the above-mentioned special points.
Then in  case (i), the array  $\left\{ \Phi _{i_{1}..i_{p}}\right\} $
that is independent of $r$ must at every $\mathbf{r}$ be invariant to the
orientation of the pair $\left( \mathbf{e}_{r},\mathbf{e}_{\theta }\right) ,$
i.e.\  ${\pmb\Phi }$ must be  transversely isotropic about $Z\parallel \mathbf{e}_{z}$;  
in  case (ii), $\left\{ \Phi _{i_{1}..i_{p}}\right\} $ that is
independent of $\theta $ must at every $\mathbf{r}$ be invariant to the
orientation of the pair $\left( \mathbf{e}_{\theta },\mathbf{e}_{\phi
}\right) $, i.e.\  ${\pmb\Phi }$ must be  transversely isotropic about $\mathbf{e}_{r};$ in  
case (iii), $\left\{ \Phi _{i_{1}..i_{p}}\right\} $ that is independent of $r
$ must at every $\mathbf{r}$ be invariant to the orientation of the
spherical frame $\left( \mathbf{e}_{r},\mathbf{e}_{\theta },\mathbf{e}_{\phi
}\right) $, i.e.\  ${\pmb\Phi }$ must be  isotropic.  
These restrictions on cylindrical or spherical
anisotropy, which result from eliminating a singularity that exists only at
isolated (axial or origin) points, are neither immanent nor  physically binding  
and can certainly be circumvented formally, say, by assuming a small 
 cavity surrounding each point. However, in the spherical case it is physically reasonable to single out  the class of 
spherically anisotropic materials  invariant with respect to any
orientation of the polar 
axis which  may thus be called  
materials with \textit{complete }spherical anisotropy. 
The remedy for  case  (ii) described above must then be provided at every $\mathbf{r}$.  Hence, material 
tensors in a body with complete spherical anisotropy can only be either
uniform or radially inhomogeneous and, unless isotropic, they must be
transversely isotropic about $\mathbf{e}_{r}$. They are well defined
everywhere except a single origin point $O$ where (iii) needs to be addressed.

\section{Governing equations}  \label{sec2}

\subsection{Elastodynamic equations}

The dynamic equilibrium vector equation for a linearly elastic material when
expressed in spherical coordinates is 
\begin{align}
& r^{-2}(r^{2}\mathbf{t}_{r})_{,\,r}+(r\sin \theta )^{-1}\big[(\sin \theta 
\mathbf{t}_{\theta })_{,\theta }+\mathbf{t}_{\phi ,\,\phi }+\sin \theta 
\mathbf{K}\mathbf{t}_{\theta }+\mathbf{H}\mathbf{t}_{\phi }\big]=\rho 
\ddot{\mathbf{u}}  \label{5} \\
& \quad \text{with }\ \ \ \mathbf{K}=%
\begin{pmatrix}
{0} & -1 & 0 \\ 
1 & 0 & 0 \\ 
0 & 0 & 0%
\end{pmatrix}%
\ \ \left( =-\mathbf{K}^{T}\right) ,\quad \mathbf{H}=%
\begin{pmatrix}
{0} & 0 & -\sin \theta \\ 
0 & 0 & -\cos \theta \\ 
\sin \theta & \cos \theta & 0%
\end{pmatrix}%
\ \ \left( =-\mathbf{H}^{T}\right) .  \notag
\end{align}%
Here $\rho =\rho ({\mathbf{x}})$ is the mass density, ${\mathbf{u}}={\mathbf{%
u}}({\mathbf{x}},t)$ the displacement and $\mathbf{t}_{i}=\mathbf{t}_{i}({%
\mathbf{x}},t)=\mathbf{e}_{i}^{T}\pmb{\sigma}$ ($i=r,\theta ,\phi $) the
traction vectors defined by the elements of stress $\pmb{\sigma}({\mathbf{x}}%
,t)$ in the orthonormal basis $\left( \mathbf{e}_{r},\mathbf{e}_{\theta },%
\mathbf{e}_{\phi }\right) $ of the spherical coordinates $\left( r,\theta
,\phi \right) $.  
The  left member in \eqref{5}, 
div$\, \pmb{\sigma}=\sum_{i}\nabla \cdot(\mathbf{e}_{i}\otimes \mathbf{t}_{i})$,  follows using   \eqref{29=}$_1$ plus the identities
$\frac{\partial \mathbf{a}}{\partial \theta} = \mathbf{a}_{,\theta } +\mathbf{K}\mathbf{a}$, 
$\frac{\partial \mathbf{a}}{\partial \phi} = \mathbf{a}_{,\phi } +\mathbf{H}\mathbf{a}$, 
where the comma suffix notation indicates 
partial differentiation of components only: $\mathbf{a}_{,\varphi }\equiv \sum_{i}a_{i,\varphi }\mathbf{e}_{i}$ for $\varphi =\theta ,\phi $. 
The stress
elements in the basis $\left( \mathbf{e}_{r},\mathbf{e}_{\theta },\mathbf{e}%
_{\phi }\right) $ are $\sigma _{ij}=c_{ijkl}\varepsilon _{kl}$ where $%
c_{ijkl}=c_{ijkl}({\mathbf{x}})$ are the components of the spherically
anisotropic elastic tensor, $\varepsilon _{kl}$
are the components  of the strain ${\pmb \varepsilon} ({\mathbf{x}},t) =\tfrac 12 
\big[ {\pmb \nabla} \mathbf{u} +( {\pmb \nabla} \mathbf{u})^T \big]$, with  summation on
repeated indices assumed and $T$ for transpose. The traction vectors can therefore be written
as 
\begin{subequations}
\label{-1}
\begin{align}
& 
\begin{pmatrix}
\mathbf{t}_{r} \\ 
\\ 
\mathbf{t}_{\theta } \\ 
\\ 
\mathbf{t}_{\phi }%
\end{pmatrix}%
=%
\begin{pmatrix}
\mathbf{Q} & \mathbf{R} & \mathbf{P} \\ 
&  &  \\ 
\mathbf{R}^{T} & \mathbf{T} & \mathbf{S} \\ 
&  &  \\ 
\mathbf{P}^{T} & \mathbf{S}^{T} & \mathbf{M}%
\end{pmatrix}%
\begin{pmatrix}
\mathbf{u}_{,\,r} \\ 
\\ 
r^{-1}(\mathbf{u},_{\,\theta }+\mathbf{K}\mathbf{u}) \\ 
\\ 
(r\sin \theta )^{-1}(\mathbf{u},_{\,\phi }+\mathbf{H}\mathbf{u})%
\end{pmatrix}%
,  \label{-1a} \\
& 
\begin{pmatrix}
\mathbf{Q} & \mathbf{R} & \mathbf{P} \\ 
&  &  \\ 
\mathbf{R}^{T} & \mathbf{T} & \mathbf{S} \\ 
&  &  \\ 
\mathbf{P}^{T} & \mathbf{S}^{T} & \mathbf{M}%
\end{pmatrix}%
=%
\begin{pmatrix}
(e_{r}e_{r}) & (e_{r}e_{\theta }) & (e_{r}e_{\phi }) \\ 
&  &  \\ 
(e_{\theta }e_r) & (e_{\theta }e_{\theta }) & (e_{\theta }e_{\phi }) \\ 
&  &  \\ 
(e_{\phi }e_{r}) & (e_{\phi }e_{\theta }) & (e_{\phi }e_{\phi })   
\end{pmatrix}%
,
\end{align}%
where, in the notation of \cite{Lothe76}, the matrix $\left( ab\right) $ has
components $\left( ab\right) _{jk}=a_{i}c_{ijkl}b_{l}$ for arbitrary vectors 
$\mathbf{a}$ and $\mathbf{b}$.

\subsection{Vector spherical harmonics}

Our objective is to develop separation of variables vector solutions in
the form $\mathbf{v}(r,\theta ,\phi )=\sum_{\mathbf{A}}V_{\mathbf{A}}(r)%
\mathbf{A}(\theta ,\phi )$ where the three vectors $\mathbf{A}\left( \mathbf{%
e}_{r}\right) $ are independent of $r$ and provide a complete basis for
representing vectorial functions of the spherical angles. Vector spherical
harmonics are one such set of functions.

It is useful to first introduce the angular parts $\mathbf{D}$ and $%
\mathbf{D\cdot D}\equiv D^{2}$ of the vector differential operators $\mathbf{%
\nabla }$ and $\mathbf{\nabla \cdot \nabla }\equiv \nabla ^{2}\left( \equiv
\Delta \right) $ in spherical coordinates: 
\end{subequations}
\begin{align}
\mathbf{\nabla } &= \mathbf{\mathbf{e}}_{r}\frac{\partial }{%
\partial r}+\frac{1}{r}\mathbf{D\ \ }\left( \mathbf{\Rightarrow D}=\mathbf{e}%
_{\theta }\frac{\partial }{\partial \theta }+\frac{\mathbf{e}_{\phi }}{\sin
\theta }\frac{\partial }{\partial \phi }\right) ,  
\notag \\
\Delta f( \mathbf{r} ) &=\frac{1}{r^{2}}\frac{\partial }{\partial
r}\left( r^{2}\frac{\partial f}{\partial r}\right) +\frac{1}{r^{2}}D^{2}f\ 
\mathbf{,}  \label{29=} \\
\Delta \mathbf{f}( \mathbf{r} ) &=\sum\nolimits_{i}\left[ \mathbf{%
e}_{i}\frac{1}{r^{2}}\frac{\partial }{\partial r}\left( r^{2}\frac{\partial
f_{i}}{\partial r}\right) \right] +\frac{1}{r^{2}}D^{2}\mathbf{f,}  
\notag
\end{align}%
so that $\Delta f=r^{-2}D^{2}f$ and $\Delta \mathbf{f}=r^{-2}D^{2}\mathbf{f}$
for pure angular functions $f(\mathbf{e}_{r})$ and $\mathbf{f}(\mathbf{e}%
_{r})$. Another ingredient is the set of (scalar) spherical harmonics $%
Y_{n}^{m}$ of polar order $n$ and azimuthal order $m$, for which there are
several slightly different notations in use.\ Following \cite[p. 64]{Martin06}%
, let%
\begin{equation}
Y_{n}^{m}(\mathbf{e}_{r})\equiv
Y_{n}^{me}+iY_{n}^{mo}=A_{n}^{m}P_{n}^{m}(\cos \theta )e^{im\phi }\ \ 
\mathrm{with}~A_{n}^{m}=(-1)^{m}\sqrt{\frac{(2n+1)}{4\pi }\frac{(n-m)!}{%
(n+m)!}} ,  \label{in_33}
\end{equation}%
where $P_{n}^{m}(\cos \theta )$ are the associated Legendre functions of the
first kind. The functions $Y_{n}^{m}(\mathbf{e}_{r})$ satisfy the equation 
\begin{equation}
\frac{1}{\sin \theta }\frac{\partial }{\partial \theta }\left( \sin \theta 
\frac{\partial Y_{n}^{m}}{\partial \theta }\right) +\frac{1}{\sin ^{2}\theta 
}\frac{\partial ^{2}Y_{n}^{m}}{\partial \phi ^{2}}\left( \equiv
D^{2}Y_{n}^{m}\right) =-\lambda ^{2}Y_{n}^{m}\ \ \mathrm{with}~\lambda
=[n(n+1)]^{1/2}.  \label{33}
\end{equation}

In these terms, the vector spherical harmonics are defined as 
\begin{align}
{\mathbf P}_{mn}(\mathbf{e}_{r})& =\mathbf{e}_{r}Y_{n}^{m}(\mathbf{e}_{r}), 
\notag  \label{in_01} \\
{\mathbf B}_{mn}(\mathbf{e}_{r})& =\lambda ^{-1}{\pmb D}Y_{n}^{m}(\mathbf{e}%
_{r}), \\
{\mathbf C}_{mn}(\mathbf{e}_{r})& ={\mathbf B}_{mn}(\mathbf{e}_{r})\times \mathbf{e%
}_{r}  \notag
\end{align}
(see \cite[\S3.17]{Martin06} for a literature review). 
The vector harmonics are pointwise orthogonal 
\begin{equation}
{\mathbf P}_{mn}\cdot {\mathbf B}_{mn}={\mathbf B}_{mn}\cdot {\mathbf C}_{mn}={\mathbf C}%
_{mn}\cdot {\mathbf P}_{mn}=0,  \label{3==3}
\end{equation}%
and  orthonormal when integrated 
by $\operatorname{d}\Omega = \sin \theta \operatorname{d}\theta \operatorname{d}\phi$: 
\begin{align}
\int_{\Omega }\operatorname{d}\Omega \, {\mathbf P}_{mn}\cdot {\mathbf B}_{\mu \nu }^{\ast }
&=\int_{\Omega }\operatorname{d}\Omega \, {\mathbf B}_{mn}\cdot {\mathbf C}_{\mu \nu }^{\ast
}=\int_{\Omega }\operatorname{d}\Omega \, {\mathbf C}_{mn}\cdot {\mathbf P}_{\mu \nu }^{\ast
}=0,  \label{-332} \\
\int_{\Omega }\operatorname{d}\Omega \, {\mathbf P}_{mn}\cdot {\mathbf P}_{\mu \nu }^{\ast }
&=\int_{\Omega }\operatorname{d}\Omega \, {\mathbf B}_{mn}\cdot {\mathbf B}_{\mu \nu }^{\ast
}=\int_{\Omega }\operatorname{d}\Omega \, {\mathbf C}_{mn}\cdot {\mathbf C}_{\mu \nu }^{\ast
}=\delta _{m\mu }\delta _{n\nu } , \notag
\end{align}%
where the latter two identities hold at $n,\nu \ne 0$, see \S\ref{sec5}(\ref{n}).   Using \cite[p.\ 1901]{MorseII}, the vector spherical harmonics can be shown to satisfy the
following identities where $f=f(r)$ and  indices $m,n$ are suppressed:%
\begin{align}
{\pmb \nabla} \cdot f{\mathbf P} &=r^{-2}{(r^{2}f)^{\prime }}Y,
\quad  {\pmb \nabla} \cdot f{\mathbf B}%
=-r^{-1}f{\lambda }Y,
\quad {\pmb \nabla} \cdot f{\mathbf C}=0;  
\notag \\
\Delta \left( f\mathbf{A}\right) &=\left( \Delta f\right) \mathbf{A}%
+r^{-2}f D^{2}\mathbf{A}\ \ \mathrm{where}\ \mathbf{A\equiv P},~\mathbf{B},~%
\mathbf{C}\ \mathrm{and}  \label{34} 
\\
D^{2}\mathbf{P} &= 2\lambda \mathbf{B}-\left( \lambda ^{2}+2\right) \mathbf{P%
},\ D^{2}\mathbf{B}=  2 \lambda  \mathbf{P}-\lambda^2 \mathbf{B}  ,\
D^{2}\mathbf{C}=-\lambda ^{2}\mathbf{C.}  \notag
\end{align}

\section{The Stroh formalism in spherical coordinates}

\label{sec3}

We consider time harmonic motion with the time dependence $e^{-i\omega t}$
omitted but understood. It is useful to re-write the equilibrium equation %
\eqref{5} in the form distinguishing terms with radial and angular
derivatives, namely: 
\begin{subequations}
\label{-21}
\begin{align}
& (r^{2}\mathbf{t}_{r})_{,r}+r\,\pmb{\tau }=-r^{2}\rho \omega ^{2}\mathbf{u},
\label{-21a} \\
& \mathrm{with}\ \pmb{\tau}\equiv (\sin \theta )^{-1}\big[(\sin \theta 
\mathbf{t}_{\theta })_{,\theta }+\mathbf{t}_{\phi ,\phi }+\sin \theta 
\mathbf{Kt}_{\theta }+\mathbf{Ht}_{\phi }\big],
\end{align}%
where $\mathbf{t}_{i}$ are defined by \eqref{-1a}. In the following, the
density and elastic coefficients are assumed to be radially inhomogeneous.
Suppress the indices $m,n$, denote $\mathbf{A}\left( \mathbf{e}_{r}\right) 
\mathbf{\equiv P},\mathbf{B},\mathbf{C}$ and let 
\end{subequations}
\begin{subequations}
\label{22}
\begin{align}
& \mathbf{u}=\sum\nolimits_{\mathbf{A}}U_{\mathbf{A}}\mathbf{A},\quad 
\mathbf{t}_{r}=\sum\nolimits_{\mathbf{A}}T_{\mathbf{A}}\mathbf{A},\quad %
\pmb{\tau}=\sum\nolimits_{\mathbf{A}}\Gamma _{\mathbf{A}}\mathbf{\mathbf{A;}}
\label{22a} \\
& \mathbf{U}=\left( U_{\mathbf{P}},U_{\mathbf{B}},U_{\mathbf{C}}\right)
^{T},\ \ \mathbf{T}=\left( T_{\mathbf{P}},T_{\mathbf{B}},T_{\mathbf{C}%
}\right) ^{T},\ \ \pmb{\Gamma }=\left( \Gamma _{\mathbf{P}},\Gamma _{\mathbf{%
B}},\Gamma _{\mathbf{C}}\right) ^{T}.  \label{22b}
\end{align}%
\end{subequations}
Our objective in this paper is two-fold:  first  to find the most general
symmetry which admits of the separation of variables solution $\mathbf{u}%
=\sum\nolimits_{\mathbf{A}}U_{\mathbf{A}}\left( r\right) \mathbf{A}$; and second,  
to obtain these separation of variables solutions when they are possible.

In view of eqs.\ \eqref{-21}-\eqref{22}, the first problem implies answering the
following question: given the \textit{ansatz} $\mathbf{U}=\mathbf{U}\left(
r\right) $, what symmetry yields simultaneous conditions $\mathbf{T}=\mathbf{%
T}\left( r\right) $ and $\pmb{\Gamma }=\pmb{\Gamma }\left( r\right) $?
Direct calculation of $T_{\mathbf{A}}=\mathbf{t}_{r}\cdot \mathbf{A}$ from (%
\ref{-1a}) with $\mathbf{u}=\sum\nolimits_{\mathbf{A}}U_{\mathbf{A}}\left(
r\right) \mathbf{A}$ shows that $\mathbf{T}=\mathbf{T}\left( r\right) $
holds for tetragonal, cubic and transversely-isotropic symmetries if their 
principal axes are parallel to $\mathbf{e}_{r}$, but it is
invalid for any other cases including the above symmetries with non-radial
principal axes, the trigonal symmetry and certainly any lower symmetries.
Note that the other traction vectors $\mathbf{t}_{\theta }$ and $\mathbf{t}%
_{\phi }$ do not admit a similar expansion in $\mathbf{A}$ with scalar
coefficients depending on $r$ even if the material is isotropic and uniform;
however, this is not so for their combination $\pmb{\tau}$. Calculation of $%
\Gamma _{\mathbf{A}}=\pmb{\tau}\cdot \mathbf{A}$ shows that $\pmb{\Gamma
}=\pmb{\Gamma }\left( r\right) $ is possible, but it holds only for  
transverse isotropy with the principal axis along $\mathbf{e}_{r}$. Thus,
altogether the separation of variables solution $\mathbf{u}=\sum\nolimits_{%
\mathbf{A}}U_{\mathbf{A}}\left( r\right) \mathbf{A}$ is ensured only in the
presence of transverse isotropy about $\mathbf{e}_{r}$; otherwise the
coefficients $U_{\mathbf{A}}$ must depend on the spherical angles (apart from  the theoretical possibility of coincidental equalities between material
constants that is of no practical interest).

Regarding the second problem, it is solved by deriving the first-order
ordinary differential system in $r$ (solutions of such ODS are standard).
This is done in detail in the next section. For now we simply formulate the
overall result.

\begin{theorem}\label{-1-1}
Time harmonic solutions of the equations of linear elasticity in a
spherically anisotropic radially inhomogeneous body with $\rho \left(
r\right) $ and $c_{ijkl}\left( r\right) $ admit of separation of variables
using the vector spherical harmonic functions only if the material is
transversely isotropic about $\mathbf{e}_{r}$. In this case, the separable
solution for the displacement and radial-traction vectors is in the form $%
\mathbf{u}=\sum\nolimits_{\mathbf{A}}U_{\mathbf{A}}(r)\mathbf{A}$,\ \ $%
\mathbf{t}_{r}=\sum\nolimits_{\mathbf{A}}T_{\mathbf{A}}(r)\mathbf{A}$ (with $%
\mathbf{A\equiv P,~B,~C}$ and the indices $m$, $n$ omitted). The amplitudes $%
\mathbf{U}\left( r\right) =\left( U_{\mathbf{P}},U_{\mathbf{B}},U_{\mathbf{C}%
}\right) ^{T}$\ and $\mathbf{T}\left( r\right) =\left( T_{\mathbf{P}},T_{%
\mathbf{B}},T_{\mathbf{C}}\right) ^{T}$ are defined by the Stroh-like  ODS 
\begin{subequations}\label{00}
\begin{equation}
\pmb{\eta }^{\prime }=\frac{i\mathbf{G}}{r^{2}}\pmb{\eta }\mathrm{\
with\ }\pmb{\eta }\left( r\right) =%
\begin{pmatrix}
\mathbf{U} \\ 
ir^{2}\mathbf{T}%
\end{pmatrix}%
,\ \mathbf{G}\left( r\right) =%
\begin{pmatrix}
ir\mathcal{T}^{-1}\mathcal{R}^{T} & -\mathcal{T}^{-1} \\ 
r^{2}\left( \mathcal{Q-RT}^{-1}\mathcal{R}^{T}\right) - r^4\rho \omega ^{2}%
\mathbf{I} & -ir{\mathcal{RT}^{-1}}%
\end{pmatrix}%
,  \label{24}
\end{equation}%
where $^{\prime }=\mathrm{d}/\mathrm{d}r$, 
\begin{align}
\mathcal{T}& =\mathrm{diag}\left[ c_{11},~c_{66},~c_{66}\right] ,\  \ \mathcal{R%
}=%
\begin{pmatrix}
2c_{12} & \lambda c_{66} & 0 \\ 
-\lambda c_{12} & -c_{66} & 0 \\ 
0 & 0 & -c_{66}%
\end{pmatrix}%
,\   \label{25a} \\
\mathcal{Q}& = 
\begin{pmatrix}
\lambda ^{2}c_{66}+4\left( c_{22}-c_{44}\right) & 
 \lambda (2c_{44} -2c_{22} -c_{66}    )  & 
0 \\ 
 \lambda (2c_{44} -2c_{22} -c_{66}    ) & \lambda ^{2}c_{22}+ c_{66}-2c_{44} & 0 \\ 
0 & 0 & \lambda ^{2}c_{44}+ c_{66}-2c_{44}
\end{pmatrix}%
=\mathcal{Q}^{T},  \label{25b}
\end{align}%
and  $\lambda =[n(n+1)]^{1/2}$. For real
material parameters (and real $\omega ^{2}$), $\mathbf{G}=\mathbb{T}\mathbf{G%
}^{+}\mathbb{T}$\textbf{\ }where $^{+}$ means Hermitian conjugation and $%
\mathbb{T}$ is a matrix with zero diagonal and identity off-diagonal 3$%
\times $3 blocks.
\end{subequations}
\end{theorem}

\section{Derivation of the Stroh-like ODS for spherical transverse isotropy}\label{sec4}

\subsection{Elastic coefficients and stress}

Having established that the separation of variables solution via the use of
harmonics works only for transverse isotropy, our purpose here is to obtain
the explicit form of the ODS which is stated by Theorem \ref{-1-1}. The
derivation proceeds by splitting the problem into the isotropic and
anisotropic parts based on the standard form of the transversely isotropic
elastic coefficients $c_{ijkl}\left( r\right) $ \cite{fed}: 
\begin{align}
c_{ijkl}& =c_{ijkl}^{(\text{iso})}+c_{ijkl}^{(\text{anis})}:\ \ c_{ijkl}^{(%
\text{iso})}=c_{23}\delta _{ij}\delta _{kl}+c_{44}\left( \delta _{ik}\delta
_{jl}+\delta _{il}\delta _{jk}\right) ,  \notag  \label{32} \\
c_{ijkl}^{(\text{anis})}& =a_{1}\left( \delta _{ik}\delta _{j1}\delta
_{l1}+\delta _{il}\delta _{j1}\delta _{k1}+\delta _{jk}\delta _{i1}\delta
_{l1}+\delta _{jl}\delta _{i1}\delta _{k1}\right) \\
& \quad +a_{2}\,\delta _{i1}\delta _{j1}\delta _{k1}\delta _{l1}+a_{3}\left(
\delta _{ij}\delta _{k1}\delta _{l1}+\delta _{kl}\delta _{i1}\delta
_{j1}\right) \ \ \mathrm{with}  \notag \\
a_{1}=c_{66}-c_{44},& \ a_{2}=c_{11}+c_{22}-2c_{12}-4c_{66},\
a_{3}=c_{12}-c_{23},\ c_{22}=c_{23}+2c_{44}.  \notag
\end{align}%
This partitions   stress as 
\begin{align}
{\pmb\sigma }& ={\pmb\sigma }^{(\text{iso})}+{\pmb\sigma }^{(\text{anis}%
)},\quad {\pmb\sigma }^{(\text{iso})}=c_{23}\left( \mathbf{\nabla \cdot u}%
\right) \mathbf{I}+2c_{44}{\pmb\varepsilon },  \notag  \label{s11} \\
{\pmb\sigma }^{(\text{anis})}& =2a_{1}\left( \mathbf{e}_{r}\otimes {\pmb%
\gamma }_{r}+{\pmb\gamma }_{r}\otimes \mathbf{\ e}_{r}\right) +a_{2}u_{r,r}%
\mathbf{e}_{r}\otimes \mathbf{e}_{r}+a_{3}\left[ u_{r,r}\mathbf{I}+\left( 
\mathbf{\nabla \cdot u}\right) \mathbf{e}_{r}\otimes \mathbf{e}_{r}\right] \\
& \mathrm{where\ }\ {\pmb\gamma }_{r}=\mathbf{e}_{r}{\pmb\varepsilon }=\frac{%
1}{2}\big(\mathbf{u}_{,r}+\mathbf{\nabla }u_{r}-r^{-1}\mathbf{u}_{\perp }%
\big)\ \ \mathrm{with\ }\mathbf{u}_{\perp }\equiv u_{\theta }\mathbf{e}%
_{\theta }+u_{\phi }\mathbf{e}_{\phi }.  \notag
\end{align}%
Another useful split occurs due to uncoupling of the shear wave motion with $%
\mathbf{u}$ polarized along $\mathbf{C}$. We begin with this observation and
subsequently examine solutions with $\mathbf{u}$ spanned by $\mathbf{P}$ and 
$\mathbf{B}$.

\subsection{Uncoupled SH solution}

Keeping the conventional term 'shear-horizontal' (SH) for the waves with $%
\mathbf{u}$ parallel to $\mathbf{C}$, assume that  
\begin{equation}
\mathbf{u}_{SH}=U_{C}(r)\,\mathbf{C}.  \label{-340}
\end{equation}%
Inserting the \textit{ansatz} \eqref{-340} into \eqref{s11} readily
determines the SH radial traction as%
\begin{equation}
\mathbf{t}_{r}=T_{C}\mathbf{C}\mathrm{\ with}\ \ T_{C}=c_{66}(U_{C}^{\prime
}-r^{-1}U_{C})  \label{3-9}
\end{equation}%
(note that the SH tractions $\mathbf{t}_{\theta }$ and $\mathbf{t}_{\phi }$
are not aligned with $\mathbf{C}$). Applying the divergence identities \eqref{34} to the isotropic and anisotropic parts of the stress gives the equalities 
\begin{align}
\operatorname{div}\pmb{\sigma }_{SH}^{(\text{iso})}& =\left[ c_{44}^{\prime }\big(%
U_{C}^{\prime }-r^{-1}U_{C}\big)+c_{44}\big(\Delta U_{C}-\lambda
^{2}r^{-2}U_{C}\big)\right] \mathbf{C},  \label{-32} \\
\operatorname{div}\pmb{\sigma }_{SH}^{(\text{anis})}& =\left[ a_{1}^{\prime }\big(%
U_{C}^{\prime }-r^{-1}U_{C}\big)\mathbf{C}+a_{1}\big(\Delta
U_{C}-2r^{-2}U_{C}\big)\right] \mathbf{C}.  \notag
\end{align}%
Adding them leads to the equation of SH motion $\operatorname{div}\pmb{\sigma }%
_{SH}=-\rho \omega ^{2}\mathbf{u}_{SH}$ in the form%
\begin{equation}
\big(r^{2}c_{66}U_{C}^{\prime }\big)^{\prime }+\big(r^{2}\rho \omega
^{2}-\lambda ^{2}c_{44}+2(c_{44}-c_{66})-rc_{66}^{\prime }\big)U_{C}=0.
\label{5-1}
\end{equation}%
The latter can be recast, using \eqref{3-9}, as%
\begin{equation}
\big(r^{2}c_{66}U_{C}^{\prime }\big)^{\prime }-\big(2c_{66}+rc_{66}^{\prime }%
\big)U_{C}=\left( r^{2}T_{C}\right) ^{\prime }+rT_{c}.  \label{-33}
\end{equation}%
Thus \eqref{3-9} is the first and \eqref{-33} the second equations of the
following ODS for SH waves:%
\begin{equation}
\begin{pmatrix}
U_{C} \\ 
ir^{2}T_{C}%
\end{pmatrix}%
^{\prime }=i%
\begin{pmatrix}
-ir^{-1} & -(r^{2}c_{66})^{-1} \\ 
(\lambda ^{2}-2)c_{44}-r^{2}\rho \omega ^{2} & ir^{-1}%
\end{pmatrix}%
\begin{pmatrix}
U_{C} \\ 
ir^{2}T_{C}%
\end{pmatrix}%
.  \label{404}
\end{equation}%
This is  clearly the same as the pair of equations in the third and sixth
rows of system \eqref{24}.

\subsection{In-plane problem}

We now consider $\mathbf{U}=\sum\nolimits_{\mathbf{A}}U_{\mathbf{A}}\mathbf{A}
$ and $\mathbf{T}=\sum\nolimits_{\mathbf{A}}T_{\mathbf{A}}\mathbf{A}$ where $%
\mathbf{A=P,B}$ only, in which sense this case may be referred to as the
'in-plane' problem. It implies restricting attention to the upper 2$\times $%
2 block of the matrices in \eqref{00}.  

\paragraph{Isotropic part}

Assume 
\begin{equation}
\mathbf{u}_{in-plane}=\sum\nolimits_{\mathbf{A}}U_{\mathbf{A}}(r)\mathbf{A }
\ \ 
\mathrm{with\ }\mathbf{A=P,B.}  \label{-341}
\end{equation}
The equation for $\mathbf{t}_{r}=\mathbf{e}_{r}\pmb{\sigma
}$ readily follows from $\pmb{\sigma }^{(\text{iso})}$ in \eqref{s11}$_{1}$
to yield the isotropic part of the first equation of the in-plane ODS as
\begin{equation}
r\mathbf{T}_{\text{iso}}=r\mathcal{T}_{\text{iso}}\mathbf{U}^{\prime }+%
\mathcal{R}_{\text{iso}}^{T}\mathbf{U\ }\mathrm{with\ }\mathcal{T}_{\text{iso%
}}=\left( 
\begin{array}{cc}
c_{23} & 0 \\ 
0 & c_{44}%
\end{array}%
\right) ,\ \mathcal{R}_{\text{iso}}^{T}=\left( 
\begin{array}{cc}
2c_{23} & -\lambda c_{23} \\ 
\lambda c_{44} & -c_{44}%
\end{array}%
\right) .  \label{s14}
\end{equation}%
In order to formulate the second equation, which is based on the equilibrium
condition $\operatorname{div}{\pmb\sigma }=-\rho \omega ^{2}\mathbf{u}$, it is
convenient to use the following identity for the radially inhomogeneous
medium:%
\begin{equation}
\left( r^{2}\mathbf{t}_{r}\right) _{,r}+r^{2}\rho \omega ^{2}\mathbf{u}=%
\left[ \left( r^{2}\mathbf{t}_{r}\right) _{,r}-r^{2}\operatorname{div}{\pmb\sigma }%
\right] _{\mathrm{homo}},  \label{s15}
\end{equation}%
where the 'homogeneous' term on the right hand side is understood to be
evaluated as if the elastic moduli are independent of $r$, i.e., derivatives
of $c_{ijkl}(r)$ are ignored. This reduces the task to concentrating on the
right member in \eqref{s15}, namely, to expanding $\operatorname{div}\pmb{\sigma }_{%
\mathrm{homo}}$ in vector spherical harmonics given that $\mathbf{u}=\sum_{%
\mathbf{A}}U_{\mathbf{A}}\mathbf{A}$: 
\begin{equation}
\operatorname{div}\pmb{\sigma }_{\mathrm{homo}}=\sum\nolimits_{\mathbf{A}}F_{\mathbf{%
A}}\mathbf{A}  
\ \ 
\mathrm{with\ }\mathbf{A=P,B.}
\label{s16}
\end{equation}%
Using $\operatorname{div}\pmb{\sigma }_{\mathrm{homo}}^{(\text{iso})}=\left(
c_{23}+c_{44}\right) \mathbf{\nabla }\left( \mathbf{\nabla \cdot u}\right)
+c_{44}\mathbf{\nabla }^{2}\mathbf{u}$ and the identities \eqref{34} yields
the isotropic part $\mathbf{F}_{\text{iso}}$ of the vector $\mathbf{F}%
=\left( F_{\mathbf{P}},F_{\mathbf{B}}\right) ^{T}$ as 
\begin{equation}
\mathbf{F}_{\text{iso}}=%
\begin{pmatrix}
c_{22}\Delta -\frac{1}{r^{2}}\left( 2c_{22}+\lambda ^{2}c_{44}\right) & 
\frac{\lambda }{r}\left[ \left( c_{44}-c_{22}\right) \frac{\mathrm{d}}{%
\mathrm{d}r}+\frac{1}{r}\left( c_{22}+c_{44}\right) \right] \\ 
\frac{\lambda }{r}\left[ \left( c_{22}-c_{44}\right) \frac{\mathrm{d}}{%
\mathrm{d}r}+\frac{1}{r}c_{22}\right] & c_{44}-\frac{\lambda ^{2}}{r^{2}}%
c_{22}%
\end{pmatrix}%
\mathbf{U}.  \label{s18}
\end{equation}%
Combining (\ref{s18}) with (\ref{s14}) and substituting into (\ref{s15})
gives%
\begin{equation}
\left( r^{2}\mathbf{T}_{\text{iso}}\right) ^{\prime }=r\mathcal{R}_{\text{iso%
}}\mathbf{U}^{\prime }+\left( \mathcal{Q}_{\text{iso}}-r^{2}\rho \omega ^{2}%
\mathbf{I}\right) \mathbf{U}\ \mathrm{with}\ \mathcal{Q}_{\text{iso}}=%
\begin{pmatrix}
4\left( c_{22}-c_{44}\right) +\lambda ^{2}c_{44} & \lambda \left(
c_{44}-2c_{22}\right) \\ 
\lambda \left( c_{44}-2c_{22}\right) & -c_{44}+\lambda ^{2}c_{22}%
\end{pmatrix}%
.  \label{s19}
\end{equation}

\paragraph{Anisotropic part}

From \eqref{s11}$_{2}$ and the definition of $a_{j}$ in \eqref{32}, the
anisotropic part to be added to the first equation is 
\begin{align}
& r\mathbf{T}_{\text{anis}}=r\mathcal{T}_{\text{anis}}\mathbf{U}^{\prime }+%
\mathcal{R}_{\text{anis}}^{T}\mathbf{U}\ \ \mathrm{with\ }  \label{s20} \\
& \mathcal{T}_{\text{anis}}=%
\begin{pmatrix}
c_{11}-c_{23} & 0 \\ 
0 & c_{66}-c_{44}%
\end{pmatrix}%
,\ \mathcal{R}_{\text{anis}}^{T}=%
\begin{pmatrix}
2\left( c_{12}-c_{23}\right) & \lambda \left( c_{23}-c_{12}\right) \\ 
\lambda \left( c_{66}-c_{44}\right) & c_{44}-c_{66}%
\end{pmatrix}%
.  \notag
\end{align}%
Again using \eqref{s11}$_{2}$, now  with  the identities 
\begin{align}
& \mathbf{\nabla }\left[ \left( \mathbf{\nabla \cdot u}\right) \mathbf{e}_{r}%
\mathbf{\otimes \mathbf{e}}_{r}\right] =\big(\Delta +\frac{2}{r}\frac{%
\partial }{\partial r}-\frac{2}{r^{2}}\big)u_{,r}\mathbf{e}_{r},\ \mathbf{\
\nabla }\left[ f( \mathbf{r} ) \mathbf{e}_{r}\mathbf{\otimes 
\mathbf{e}}_{r}\right] =\big(f_{,r}+\frac{2}{r}f\big)\mathbf{\mathbf{e}}_{r},
\label{s21} \\
& \mathbf{\nabla }\left[ f\left( r\right) \left( \mathbf{e}_{r}\mathbf{%
\otimes B}+\mathbf{B\otimes e}_{r}\right) \right] =f^{\prime }\mathbf{B}+f%
\mathbf{\nabla }\left( \mathbf{e}_{r}\mathbf{\otimes B}+\mathbf{B\otimes e}%
_{r}\right) =f^{\prime }\mathbf{B}+r^{-1}f\left( 3\mathbf{B-}\lambda \mathbf{%
P}\right) ,  \notag
\end{align}%
yields the anisotropic part $\mathbf{F}_{\text{anis}}$ of the vector $%
\mathbf{F}=\left( F_{\mathbf{P}},F_{\mathbf{B}}\right) ^{T}$ defined in 
(\ref{s16}): 
\begin{equation}
\mathbf{F}_{\text{anis}}=%
\begin{pmatrix}
\left( a_{2}+2a_{3}\right) \Delta +\frac{1}{r^{2}}\left(
2a_{3}-4a_{1}-\lambda ^{2}a_{1}\right) & \frac{\lambda }{r}\left[
(a_{1}-a_{3})\frac{\mathrm{d}}{\mathrm{d}r}+\frac{1}{r}\left(
3a_{1}-a_{3}\right) \right] \\ 
\frac{\lambda }{r}\left( (a_{3}-a_{1})\frac{\mathrm{d}}{\mathrm{d}r}+\frac{2%
}{r}a_{1}\right) & a_{1}\left( \Delta -\frac{2}{r^{2}}\right)%
\end{pmatrix}%
\mathbf{U}.  \label{s22}
\end{equation}%
Combining (\ref{s22}) with (\ref{s20}) and substituting the result into (\ref%
{s15}) gives 
\begin{equation}
\left( r^{2}\mathbf{T}_{\text{anis}}\right) ^{\prime }=r\mathcal{R}_{\text{%
anis}}\mathbf{U}^{\prime }+\left( \mathcal{Q}_{\text{anis}}-r^{2}\rho \omega
^{2}\mathbf{I}\right) \mathbf{U}\ \mathrm{with}\ \mathcal{Q}_{\text{anis}%
}=\left( c_{66}-c_{44}\right) 
\begin{pmatrix}
\lambda ^{2} & -\lambda \\ 
-\lambda & 1%
\end{pmatrix}%
.  \label{s23}
\end{equation}

\paragraph{Result:}

The isotropic and anisotropic parts of the in-plane solution can now be
superimposed. Adding (\ref{s20}) to (\ref{s14}) and (\ref{s23}) to (\ref{s19}%
) gives the in-plane part of 
\eqref{24}. This, together with \eqref{404}, completes the proof of Theorem \ref%
{-1-1}.

\section{Discussion}

\label{sec5}

\subsection{Radially uniform materials}\label{5.1}

Consider first the SH motion. Assuming radially uniform material
coefficients and denoting $U_{C}(r)\rightarrow u(R)$ with $R=r\omega \big(
\rho /c_{66}\big)^{1/2}$, the uncoupled equation (\ref{5-1}) becomes 
\begin{equation}
\big(R^{2}u^{\prime }\big)^{\prime }+[ R^{2}-\mu (\mu +1)]\, u=0,\quad \mu (\mu
+1)=[n(n+1)-2]\frac{c_{44}}{c_{66}}+2,  \label{5-6}
\end{equation}%
the solutions of which are spherical Bessel functions $j_{\mu }(R)$ and $%
y_{\mu }(R)$ (alternatively, Hankel functions $h_{\mu }^{(1)}(R)$, $h_{\mu }^{(2)}(R)$).  The identity \eqref{5-6}$_2$ was obtained by \cite{Hu54} for static equilibrium, and the vector function defined by \eqref{-340} with Bessel 
or Hankel  function solutions for $U_{C}(r)$
is equivalent to the vector function ${\pmb M}(r,\theta ,\phi )$ of \cite{MorseII} (see also \cite{Dassios95}).

The in-plane differential equations for displacement
and traction amplitudes may be recast as second-order differential
equations for $U_{P}$ and $U_{B}$. In the case of radially uniform
materials, these equations reduce to 
\begin{align}
\begin{pmatrix}
0 \\ 
0%
\end{pmatrix}%
& =%
\begin{pmatrix}
c_{11} & 0 \\ 
0 & c_{66}%
\end{pmatrix}%
\begin{pmatrix}
r^{2}U_{P}^{\prime } \\ 
r^{2}U_{B}^{\prime }%
\end{pmatrix}%
^{\prime }+r\lambda (c_{12}+c_{66})%
\begin{pmatrix}
0 & -1 \\ 
1 & 0%
\end{pmatrix}%
\begin{pmatrix}
U_{P}^{\prime } \\ 
U_{B}^{\prime }%
\end{pmatrix}
\notag  \label{5-8} \\
& +%
\begin{pmatrix}
\rho \omega ^{2}r^{2}-2(c_{22}+c_{23}-c_{12})-\lambda ^{2}c_{66} & \lambda
(c_{22}+c_{23}-c_{12}+c_{66}) \\ 
\lambda (c_{22}+c_{23}+2c_{66}) & \rho \omega
^{2}r^{2}+2(c_{44}-c_{66})-\lambda ^{2}c_{22}%
\end{pmatrix}%
\begin{pmatrix}
U_{P} \\ 
U_{B}%
\end{pmatrix}%
.
\end{align}%
Even though the material coefficients are constant, this coupled pair of
equations does not seem to admit an evident explicit solution in terms of
special functions and should be solved by means discussed in \S \ref{sec5}(\ref{5b}).
For isotropic materials with $c_{L}^{2}=c_{11}/\rho $ and $%
c_{T}^{2}=c_{66}/\rho $, eqs.\ \eqref{5-8} further simplify to the form 
\begin{subequations}
\label{-46}
\begin{align}
(r^{2}U_{P}^{\prime })^{\prime }+(\frac{\omega ^{2}}{c_{L}^{2}}r^{2}-\lambda
^{2}-2)U_{P}+2\lambda U_{B}-\lambda
c_{L}^{-2}(c_{L}^{2}-c_{T}^{2})[(r U_{B})^{\prime }-\lambda U_{P}]& =0,
\label{-46a} \\
(r^{2}U_{B}^{\prime })^{\prime }+(\frac{\omega ^{2}}{c_{T}^{2}}r^{2}-\lambda
^{2})U_{B}+2\lambda U_{P}+\lambda c_{T}^{-2}(c_{L}^{2}-c_{T}^{2})[\frac{1}{r}%
(r^{2}U_{P})^{\prime }-\lambda U_{B}]& =0,  \label{-46b}
\end{align}%
which leads to the known isotropic solutions. They may be elicited as
follows. Set the final term in \eqref{-46a} to zero by assuming for some as
yet unknown function $v$ that 
\end{subequations}
\begin{equation}
U_{P}=v^{\prime },\quad U_{B}=\lambda \frac{v}{r}.  \label{-47}
\end{equation}%
Substituting this into equations \eqref{-46} recasts them, respectively, as 
\begin{equation}
\big(\frac{\operatorname{d}}{\operatorname{d}r}-\frac{2}{r}\big)M_{L}v=0,\quad \frac{\lambda 
}{r}\frac{c_{L}^{2}}{c_{T}^{2}}M_{L}v=0,\quad \text{with }\ M_{L}v\equiv
(r^{2}v^{\prime })^{\prime }+(\frac{\omega ^{2}}{c_{L}^{2}}r^{2}-\lambda
^{2})v.  \label{-49}
\end{equation}%
Both equalities are satisfied if $v$ is a solution of the same spherical
Bessel equation, $M_{L}v=0$, which has solutions $j_{n}(r\omega /c_{L})$ and 
$y_{n}(r\omega /c_{L})$. Similarly, setting to zero the final term in %
\eqref{-46b} by taking%
\begin{equation}
U_{P}=\lambda \frac{w}{r},\quad U_{B}=\frac{1}{r}(rw)^{\prime },  \label{-57}
\end{equation}%
for some function $w$, we find that \eqref{-46a} and \eqref{-46b} become,
respectively, 
\begin{equation}
\frac{\lambda }{r}\frac{c_{T}^{2}}{c_{L}^{2}}M_{T}w=0,\quad 
\big(\frac{\operatorname{d}}{\operatorname{d}r}-\frac{1}{r}\big)M_{T}w=0,\quad \text{with }\ M_{T}w\equiv
(r^{2}w^{\prime })^{\prime }+(\frac{\omega ^{2}}{c_{T}^{2}}r^{2}-\lambda
^{2})w.  \label{-59}
\end{equation}%
These are satisfied if $w$ is a solution of the spherical Bessel equation of
order $n$, $M_{T}w=0$. 

The uncoupled longitudinal and transverse wave solutions \eqref{-47} and %
\eqref{-57} for the uniform and isotropic case are consistent with the
potential representation using the Helmholtz decomposition. The vector
functions of $r,\theta ,\phi $ formed from \eqref{-341} with \eqref{-47}, %
\eqref{-57} can be identified as the vector functions $\pmb L$ and $\pmb N$,
respectively, of \cite{MorseII} (see also \cite{Dassios95}).

\subsection{Radially inhomogeneous materials} \label{5b}

The first-order ODS (\ref{24}) with material coefficients depending on $r$
admits a general solution in the standard form $\pmb{\eta }(
r) =\mathbf{M}( r,r_{0}) \pmb{\eta }( r_{0})
, $ where $\pmb{\eta }( r_{0}) $ is the initial data and the
matricant $\mathbf{M}( r,r_{0}) $ may be evaluated by the Peano
series \cite{Pease} 
\begin{equation}
\mathbf{M}( r,r_{0}) =\mathbf{I+}\int_{r_{0}}^{r}\mathrm{d}x\frac{%
i\mathbf{G}( x) }{x^{2}}+\int_{r_{0}}^{r}\mathrm{d}x\frac{i%
\mathbf{G}( x) }{x^{2}}\int_{r_{0}}^{x}\mathrm{d}x_{1}\frac{i%
\mathbf{G}( x_{1}) }{x_{1}^{2}}+...  \label{+1}
\end{equation}%
The matricant solution applies for $r,r_{0}\neq 0$. The case where the
solution needs to be extended to the origin point $r=0$ requires a special
treatment based on the theory of ODS with an irregular singular point \cite{Wasow}. A similar state of affairs arises for the Stroh-like ODS in
cylindrical coordinates at the axial points $r=0$ (except that they are
regular singular points) \cite{Shuvalov03a,Norris10}.
In the following we assume $r,r_{0}\neq 0.$

By analogy with the cylindrical case, the algebraic symmetry $\mathbf{G}=%
\mathbb{T}\mathbf{G}^{+}\mathbb{T}$ (see Theorem \ref{-1-1}) yields $\mathbf{%
M}^{+}\mathbb{T}\mathbf{M}=\mathbb{T}$ and 
\begin{equation}
\frac{\mathrm{d}}{\mathrm{d}r}\left( \mathbf{\mathcal{N}}^{+}\mathbb{T}%
\mathbf{\mathcal{N}}\right) =\mathbf{0}
\quad
\mathrm{with\ \ }\mathbf{\mathcal{N}%
} (r) \equiv \left\{ \pmb{\eta }^{( \alpha )
}\right\} ,  \label{+2}
\end{equation}%
where the 6$\times $6 matrix $\mathbf{\mathcal{N}}(r) $ is
composed of any six linearly independent solutions 
$\pmb{\eta }^{( \alpha ) }( r) 
=\big( \mathbf{U}^{( \alpha ) },~ir^{2}\mathbf{T}^{( \alpha ) } \big)^{T}$, $\alpha =1..6,$
of (\ref{24}). Thus 
$\mathbf{\mathcal{N}}^{+}( r) \mathbb{T} \mathbf{\mathcal{N}}( r) $ is a constant matrix which can be
chosen to provide the partial solutions $\pmb{\eta }^{\left( \alpha
\right) }(r) $ with appropriate pointwise orthogonality in the
sense of a product (\ref{+2}) 
\cite{Shuvalov03}.

Recalling the angular indices of the vector spherical harmonics  
$\mathbf{A}_{mn}( \mathbf{e}_{r}) $$\equiv$$\mathbf{P}_{mn}$, $\mathbf{B}
_{mn}$, $\mathbf{C}_{mn}$ defines the displacement-traction modes in full as%
\begin{align}
\pmb{\eta }^{(\alpha ) }( \mathbf{r} ) 
& =\left( 
\begin{array}{c}
\mathbf{u}^{(\alpha ) } \\ 
ir^{2}\mathbf{t}_{r}^{(\alpha ) }%
\end{array}%
\right) =\sum_{n=0}^{\infty }\pmb{\eta }_{n}^{(\alpha ) }
\ \  \ (\alpha = 1\ldots 6) 
\quad 
\mathrm{ with\ }
\notag \\ 
\pmb{\eta }_{n}^{(\alpha ) }( \mathbf{r} )
 &=\left( 
\begin{array}{c}
\mathbf{u}_{n}^{(\alpha ) } \\ 
ir^{2}\mathbf{t}_{nr}^{(\alpha ) }%
\end{array}%
\right) 
=\sum_{\mathbf{A}}%
\begin{pmatrix}
U_{\mathbf{A}_{mn},n}^{(\alpha ) }(r)
\sum_{\left\vert m\right\vert <n}\mathbf{A}_{mn} \\ 
ir^{2}T_{\mathbf{A}_{mn},n}^{(\alpha ) }(r)
\sum_{\left\vert m\right\vert <n}\mathbf{A}_{mn}%
\end{pmatrix}%
,  \label{+3}
\end{align}%
where it is taken into account that $U_{\mathbf{A}_{mn},n}^{( \alpha) }$, $T_{\mathbf{A}_{mn},n}^{(\alpha ) }(r) $
are independent of $m$ (see \S \ref{sec5}(\ref{m})). There is no pointwise orthogonality
of the 'full' modes in general due to the complex conjugation in (\ref{+2})
and its absence in (\ref{3==3}). At the same time, one can make use of (\ref%
{+2}) along with the integral orthogonality of vector harmonics (\ref{-332}%
). This enables evaluation of the angular (and time-period) average of the
radial component of energy flux associated with the mode $%
\pmb{\eta }^{(\alpha ) }( \mathbf{r} ) $ as 
\begin{equation}
P_{r}^{(\alpha ) }\equiv -\int_{\Omega }\left\langle \mathbf{t}%
_{r}^{(\alpha ) }\cdot \mathbf{\dot{u}}^{(\alpha )
}\right\rangle _{t}\operatorname{d}\Omega =-\frac{\omega }{4r^{2}}\int_{\Omega }%
\pmb{\eta }^{(\alpha ) +}( \mathbf{r} ) \mathbb{%
T}\pmb{\eta }^{(\alpha ) }( \mathbf{r} ) 
\operatorname{d}\Omega . \label{+4}
\end{equation}%
 This may be further reduced   by using  (\ref%
{-332}), (\ref{+2}) and (\ref{+4}), 
\begin{equation}
P_{r}^{(\alpha ) }
=\sum_{n=0}^{\infty }P_{nr}^{( \alpha ) },\ 
\mathrm{where}\ 
P_{nr}^{(\alpha ) }
\equiv
-\int_{\Omega }\left\langle \mathbf{t}_{rn}^{(\alpha ) }\cdot 
\mathbf{\dot{u}}_{n}^{(\alpha ) }\right\rangle _{t}
\operatorname{d} \Omega 
= -\frac{\omega }{4r^{2}}\left( 2n+1\right) \pmb{\eta }_{n}^{( \alpha ) +}(r) \mathbb{T}\pmb{\eta }_{n}^{( \alpha ) }(r) .  \label{+5}
\end{equation}%
Thus the fluxes $P_{nr}^{(\alpha ) }$ carried by the modes 
$\pmb{\eta }_{n}^{(\alpha ) }( \mathbf{r} ) $
add up to give  $P_r^{(\alpha)}$ of (\ref{+4}).  According to (\ref{+2}), these fluxes do not depend on $r$ which is consistent with  energy
conservation for the assumed case of real material parameters.

\subsection{The case $\mathbf{u} ( r,\protect\theta ) $ ($m=0$)}

\label{m}

The separation of variables using vector spherical harmonics involves a
single auxiliary equation (\ref{33}) which does not depend on $m$, and 
hence Theorem \ref{-1-1} holds for any $m$ \cite{Shulga88}. 
This does not, however, preclude the 
possibility that some lower symmetry permits such separation for the
specific case of $m=0$. The derivation along the lines of \S 
\ref{sec3} shows that 'radially tetragonal' symmetry, with $\mathbf{e}_{r}$
parallel to the $4$-fold axis and with $c_{24},~c_{34}=0$, admits separation
of variables for $m=0$. The result for this tetragonal symmetry at $m=0$
amounts to replacing $2c_{44}$ by $c_{22}-c_{23}$ in all entries of $\mathcal{Q}$ in \eqref{25b}, except the term $\lambda ^{2}c_{44}$ in the last diagonal component.  
Any
other tetragonal symmetry (including that with $\mathbf{e}_{r}\parallel 4$%
-fold axis but with $c_{24}=-c_{34}\neq 0$), any trigonal symmetry and
indeed all lower symmetries prevent separation of variables even for $m=0$.

The same question may be raised concerning the SH uncoupling for $m=0$. The
above consideration implies that (\ref{404}) with $m=0$ can be extended to
'radially tetragonal' symmetry. Moreover, it can be shown that orthorhombic
symmetry aligned with the coordinate planes also leads to SH uncoupling for $%
m=0$, with the equation of motion obtained from (\ref{404}) by replacing $%
c_{66}$ with $c_{55}$. In this case  the SH modes admit the
separation of variables form $\mathbf{u}_{SH}( r,\theta ) =U_{%
\mathbf{C}}(r) \lambda ^{-1}P_{n,\theta }( \theta ) 
\mathbf{e}_{\phi }$ while the modes $\mathbf{u}( r,\theta ) $
polarized in the plane $\left\{ \mathbf{e}_{r},\mathbf{e}_{\theta }\right\} $
do not.  SH uncoupling is generally precluded for
tetragonal symmetry with $c_{24}=-c_{34}\neq 0$ and for trigonal symmetry
even if $m=0.$

\subsection{Solutions for $n=0$} \label{n}

Consider the term $\pmb{\eta }_{0}^{\left( \alpha \right) }\left( \mathbf{%
r}\right) $ with $n=0$ (and hence $m=0$) of the series (\ref{+3}). The
corresponding spherical harmonics are 
\begin{equation}
\mathbf{P}_{00}=\frac{1}{2\sqrt{\pi }}\, \mathbf{e}_{r},\ {\mathbf{B}}_{00}=%
\mathbf{0},\ {\mathbf{C}}_{00}=\mathbf{0}.  \label{d1}
\end{equation}
Vanishing of ${\mathbf{B}}_{00}$ and ${\mathbf{C}}_{00}$ may be formally
inferred from the definition \eqref{in_01} by taking the limit of $[\nu (\nu
+1)]^{-1/2}\frac{\partial ~}{\partial \theta }P_{\nu }(\cos \theta )$ as $%
\nu \rightarrow 0$. It is also consistent with another framework that
defines $\mathbf{B}_{mn}$ and $\mathbf{C}_{mn}$ without a normalization
factor $\lambda ^{-1}$.  Note however that the symmetry $\mathbf{G}=\mathbb{T}%
\mathbf{G}^{+}\mathbb{T}$ of the system matrix in \eqref{00} is not
preserved under any change of the definition \eqref{in_01} unless it implies
multiplying all three harmonics $\mathbf{A}_{mn}$ by the same factor.


By (\ref{d1}), the sextet $\pmb{\eta }_{0}^{\left( \alpha \right) }$ 
\textbf{\ }$(\alpha =1..6)$ for $n=0$ contains only two non-zero modes which
are the modes $\alpha =l1,l2$ of the longitudinal (radially polarized) wave 
\begin{equation}
\mathbf{u}=U(r)\, \mathbf{e}_{r},\quad \mathbf{t}_{r}=T(r)\, \mathbf{e}_{r},
\label{d2}
\end{equation}%
where we drop the indices used in (\ref{+3}). Equation \eqref{24} with $n=0$
provides an uncoupled system of two equations for the amplitudes $U$ and $T$: 
\begin{equation}
\begin{pmatrix}
U \\ 
ir^{2}T%
\end{pmatrix}%
^{\prime }=i%
\begin{pmatrix}
2ir^{-1}c_{12}c_{11}^{-1} & -r^{-2}c_{11}^{-1} 
\\ 
4\left( c_{22}-c_{44}-\frac{c_{12}^{2}}{c_{11}}\right) -r^{2}\rho \omega ^{2}
& -2ir^{-1}c_{12}c_{11}^{-1}%
\end{pmatrix}%
\begin{pmatrix}
U \\ 
ir^{2}T%
\end{pmatrix}%
,  \label{d3}
\end{equation}%
which reduces to a single second-order equation for $U(r)$,
\begin{equation}
\big(r^{2}c_{11}U^{\prime }\big)^{\prime }+\big(r^{2}\rho \omega
^{2}+4(c_{44}-c_{22})+2(rc_{12})^{\prime }\big)U=0.  \label{d4}
\end{equation}%
Note that the wave (\ref{d2}) exerts the tractions $\mathbf{t}_{i}=f\left(
r\right) \mathbf{e}_{i}\ $for$\ i=\theta ,\phi $,\ where$\ f=c_{12}U^{\prime
}+r^{-1}\left( c_{22}-c_{44}\right) U.$ Equations (\ref{d3}) and (\ref{d4})
are similar to the SH-wave eqs. (\ref{404}) and (\ref{-33}) formally 
taken with $n=0$. Since the wave (\ref{d2}) also implies $m=0,$ it follows
from \S 6(c) that eqs.\ (\ref{d3}), (\ref{d4}) obtained from the transversely
isotropic system \eqref{25a} and \eqref{25b}  can be extended to the tetragonal case, by
replacing $2c_{44}$ with $c_{22}-c_{23}$.  The same results can be
obtained by inserting $\mathbf{u}=U(r)\mathbf{e}_{r}$ in the initial 
 elastodynamic equations (\ref{5})-(\ref{-1}) with tetragonal symmetry. 
Note in this regard that the elastostatic solution of the form (\ref{d2}) was 
analyzed for various symmetries in \cite{Ting98,Antman87} and \cite{Antman01}.

Interestingly, replacing $\lambda ^{-1}$ in the harmonics definition %
\eqref{in_01} with some other power of $\lambda $ may lead to peculiar
solutions at $n=0$ different from (\ref{d2}). This is a problem of interest
in its own right. Consider the set of spherical harmonics
defined as
\begin{equation}
\label{227}
\tilde{\mathbf P}_{mn} =  {\mathbf P}_{mn}, 
\ \ 
\tilde{\mathbf B}_{mn} = \lambda^{-1}{\mathbf B}_{mn}, 
\ \ 
\tilde{\mathbf C}_{mn} = \lambda^{-1}{\mathbf C}_{mn}.  
\quad  
\end{equation}
The corresponding spherical harmonic functions for $n=0$ are (dropping the $m=0$ subscript) 
\begin{align}
\tilde{\mathbf P}_0  =A_0 \mathbf{e}_{r}, 
\quad 
\tilde{\mathbf B}_{0}  = - A_0 \tan \frac \theta 2 \, \mathbf{e}_\theta, 
\quad 
\tilde{\mathbf C}_{0}  =  A_0 \tan \frac \theta 2 \, \mathbf{e}_\phi, 
\ \ \text{with}\ \ A_0 = \frac 1{2\sqrt{\pi}},
\label{231}
\end{align}
which follow from \eqref{227} using first an exchange of  limits,  
\begin{equation}
\lim_{\nu \rightarrow 0} 
\frac{1}{ \nu (\nu +1) } 
\frac{\text{d} ~ }{\text{d} \theta} P_\nu (\cos \theta)  
= \frac{\text{d} ~ }{\text{d} \theta}
\lim_{\nu \rightarrow 0} 
\frac{ P_\nu (\cos \theta) -P_0 (\cos \theta) }{\nu (\nu +1) },
\label{4321}
\end{equation}
and then  Jolliffe's formula \cite{Jolliffe} to evaluate the derivative with respect to $\nu$, 
\begin{equation}
\left. \frac{\text{d} P_\nu (z) }{\text{d} \nu}\right|_{\nu = n} 
\equiv F_{n}(z) = 
- P_n (z) \ln \frac{z+1}{2} + \frac{2 }{ 2^n n !}
\frac{\text{d}^n  }{\text{d} z^n} \bigg( 
(z^2-1)^n \ln \frac{z+1}{2}
\bigg). 
\label{1234}
\end{equation}
By (\ref{231}), $\tilde{\mathbf{P}}_{0}$ corresponds to purely radial
motion, while $\tilde{\mathbf{B}}_{0}$ and $\tilde{\mathbf{C}}_{0}$
(non-zero in contrast to (\ref{d1})) represent shearing and twisting about
the polar axis, with zero at one pole and a singularity at the other. The
latter can be avoided by introducing a conical cut centered at $r=0$ of
arbitrarily small angular extent at $\theta =\pi $. This obviates the
singularity  allowing the otherwise un-normalized spherical harmonics
 $\tilde{\mathbf{B}}_{0}$ and $\tilde{\mathbf{C}}_{0}$ to satisfy the orthonormality
conditions \eqref{-332}.  
Note that the spherical harmonics $\tilde{\mathbf{B}}_{0}$ and $\tilde{%
\mathbf{C}}_{0}$ are derived from and related with Legendre polynomials of
the first kind which define the spherical harmonic $\tilde{\mathbf{P}}_{0}$,
although the Legendre polynomials of the second kind may be represented  
 as $Q_{n}(z)=\frac{1}{%
2}[F_{n}(z)-(-1)^{n}F_{n}(-z)]$ (see eq.\ \eqref{1234}) \cite{Jolliffe}.  


Based on the above, it is of interest to
consider specific  $n=0$  solutions in the form 
\begin{equation}
\mathbf{u}=\sum\nolimits_{\tilde{\mathbf{A}}_{0}}\tilde{U}_{\tilde{\mathbf{A}%
}_{0}}(r)\tilde{\mathbf{A}}_{0},\quad \mathbf{t}_{r}=\sum\nolimits_{\tilde{%
\mathbf{A}}_{0}}\widetilde{T}_{\tilde{\mathbf{A}}_{0}}(r)\tilde{\mathbf{A}}%
_{0},\ \ \tilde{\mathbf{A}}_{0}=\tilde{\mathbf{P}}_{0},\tilde{\mathbf{B}}%
_{0},\tilde{\mathbf{C}}_{0}.
\end{equation}%
The system \eqref{00} modified with respect to the
spherical harmonics $\tilde{\mathbf{A}}$ of \eqref{227} and taken for $n=0$
defines the displacement amplitudes $\tilde{U}_{\tilde{\mathbf{A}}_{0}}(r)$
by the following equations: 
\begin{subequations}
\begin{align}
\big(r^{2}c_{11}\tilde{U}_{\tilde{\mathbf{P}}_{0}}^{\prime }\big)^{\prime }+%
\big(r^{2}\rho \omega ^{2}+4(c_{44}-c_{22})+2(rc_{12})^{\prime }\big)\tilde{U%
}_{\tilde{\mathbf{P}}_{0}}& =\big[2(c_{44}-c_{22}+\frac{c_{12}^{2}}{c_{11}}%
)-rc_{66}^{\prime }\big]\tilde{U}_{\tilde{\mathbf{B}}_{0}}  
\notag \\
& \ \ +r^{2}\big[r^{-1}(c_{66}+c_{12})\tilde{U}_{\tilde{\mathbf{B}}_{0}}\big]%
^{\prime },  \label{11a} \\
\big(r^{2}c_{66}\tilde{U}_{\tilde{\mathbf{A}}_{0}}^{\prime }\big)^{\prime }+%
\big(r^{2}\rho \omega ^{2}+2(c_{44}-c_{66})-rc_{66}^{\prime }\big)\tilde{U}_{%
\tilde{\mathbf{A}}_{0}}&= 0,\ \ \tilde{\mathbf{A}}_{0}=\tilde{\mathbf{B}}%
_{0},\tilde{\mathbf{C}}_{0}.  \label{11b}
\end{align}
\end{subequations}
It is seen that eqs. \eqref{11b} for the amplitudes $\tilde{U}_{\tilde{%
\mathbf{B}}_{0}}$ and $\tilde{U}_{\tilde{\mathbf{C}}_{0}}$ are the same as
the SH wave equation \eqref{5-1} with $n=0,$ and that $\tilde{U}_{\tilde{%
\mathbf{B}}_{0}}$ provides a forcing term in eq.\ \eqref{11a} for $\tilde{U}_{%
\tilde{\mathbf{P}}_{0}}$ (note also that\ \eqref{11a} reduces to (\ref{d4})
if $\tilde{U}_{\tilde{\mathbf{B}}_{0}}=0$). Thus the SH solution $\mathbf{u}=%
\tilde{U}_{\tilde{\mathbf{C}}_{0}}\tilde{\mathbf{C}}_{0}$ $\left( \parallel 
\mathbf{e}_{\phi }\right) $ is completely uncoupled. However, the SH motion
polarized in the $\tilde{\mathbf{B}}_{0}\parallel \mathbf{e}_{\theta }$
direction drives the radial motion and this results in the coupled wave $%
\mathbf{u}=\tilde{U}_{\tilde{\mathbf{P}}_{0}}\tilde{\mathbf{P}}_{0}+\tilde{U}%
_{\tilde{\mathbf{B}}_{0}}\tilde{\mathbf{B}}_{0}.$ At the same time, non-zero
angular motion requires that the issue of singularities at $\theta =\pi $ be
resolved. Under normal circumstances, e.g. a solid sphere, or a complete
shell, this is not the case, and any angular motion is precluded, i.e. $%
\tilde{U}_{\tilde{\mathbf{B}}_{0}}=\tilde{U}_{\tilde{\mathbf{C}}_{0}}=0$ and
only radial motion $\mathbf{u}=U\left( r\right) \mathbf{e}_{r}$ occurs
(where $U$ is defined by (\ref{d4})). If the small conical cut device is
introduced, then the surface of the cone must apply a force and a moment
sufficient to maintain the dynamic tractions required of the solutions. The
magnitudes of the tractions for different types of motion depend critically
on the cone angle $\varepsilon \ll 1$. For purely longitudinal motion the
traction is independent of $\varepsilon $, i.e. O$(1)$. The $\left( \mathbf{e%
}_{r},\,\mathbf{e}_{\theta }\right) $-coupled motion requires normal traction 
 $\mathbf{t}_{\theta}$  of 
O$\big(\left. \tan^2\frac{\theta}{2}\right|_{\theta = \pi -  \varepsilon } \big)=$ 
O$(\varepsilon ^{-2})$ 
and is therefore ruled out as a
viable $n=0$ dynamic solution. In the case of the pure SH motion $\mathbf{u}=%
\tilde{U}_{\tilde{\mathbf{C}}_{0}}\tilde{\mathbf{C}}_{0},$ the normal 
traction  $\mathbf{t}_{\theta} = r^{-1}c_{44} \tan\frac{\theta}{2}\,\mathbf{u}$ 
is  O$(\varepsilon ^{-2})$,   a pure twist  in the $\mathbf{e}_{\phi }$ direction 
corresponding to a net torque about the polar axis of order unity $(r \varepsilon \times 2\pi r 
\varepsilon \times \varepsilon ^{-2})$. This suggests that pure twisting
motion may be induced in a solid sphere with a fixed polar axis by application of torque to a small
conical insert.

\section{Conclusion}

\label{sec6}

The central result of the paper, Theorem \ref{-1-1}, shows that 
spherically anisotropic radially inhomogeneous materials 
admit elastodynamic solutions $\mathbf{u}(r,\theta ,\phi )$ in a separation of 
variables form.  The angular dependence (on $\theta ,\phi $) is
described by the vector spherical harmonics while the radial dependence (on $%
r$) is separated and determined by the Stroh-like first-order ordinary
differential system that is solvable by standard means. It is proved that
such separation of variables solution is generally possible only if the
spherical anisotropy is restricted to transverse isotropy with the principal
 axis in the radial direction $\mathbf{e}_{r}$.  Transverse isotropy
about   $\mathbf{e}_{r}$   distinguishes a class of materials with   
\textit{complete} spherical anisotropy, which is a physically natural model
of spherical anisotropy since it ensures invariance of material properties
with respect to any orientation of  the polar axis.

The separable of variables solution $\mathbf{u}(r,\theta ,\phi )$ for the
transverse isotropic case does not explicitly depend on the azimuthal order $%
m$. At the same time, dependence on $m$ reveals itself in that the solution
of the form $\mathbf{u}(r,\theta )$ (i.e.\ with $m=0$) admits separation of
variables via spherical harmonics not only for transverse isotropy 
but also for lower symmetry - but only 'up
to' tetragonal with $\mathbf{e}_{r}$ along the $4$-fold axis and with $ c_{24}$, $c_{34}=0$. Note that the solutions $\mathbf{u}(r,\theta ,\phi )$ for
transverse isotropy  and $\mathbf{u}(r,\theta )$ for the above
tetragonal symmetry uncouple the shear modes parallel to the vector harmonic 
$\mathbf{C}$ ($\parallel \mathbf{e}_{\phi }$ at $m=0$) from the in-plane
modes orthogonal to $\mathbf{C}$.  Moreover,  shear modes of the form $%
\mathbf{u}(r,\theta )$ are uncoupled also for orthorhombic symmetry but not
for trigonal symmetry nor for tetragonal if $c_{24}$, $c_{34}\neq 0$.

The establishment of the Stroh format  for the elastodynamic equations in spherical coordinates opens the door for applications  to  various boundary value and scattering problems.     For instance, solutions for  acoustic and elastic wave  scattering from solid spheres and shells, which  have been limited to isotropic materials  (see \cite[\S 4.10]{Martin06}
for a review) or transversely isotropic shells with $m=0$ \cite{Hash09}, can be generated for arbitrarily layered shells and solids using standard solution techniques outlined in 
\S \ref{sec5}(\ref{5b}).    
Other possible approaches that can be explored with the Stroh formalism  include impedance matrices  for spherical shells and solids, the use of which simplifies the formulation of boundary value problems, e.g.  determining  modal frequencies,  solving radiation and scattering problems.  By analogy with the  cylindrical situation \cite{Norris10}  it should be possible to formulate a matrix Riccati ordinary differential equation for the impedance  matrix as a function of the spherical radius $r$, with a unique solution at the origin   that depends only on the elastic constants at $r=0$.   The Stroh formalism is particularly suited to solution of elastodynamic problems with forcing, e.g. from thermal expansion via laser excitation with application to non-destructive testing.  More exotic issues  could  be addressed with the Stroh system, such as modelling and simulation of fully elastic 3D 'radial wave crystals'  \cite{Torrent09}, i.e.\ shells of
radially periodic  materials that exhibit Bloch wave effects normally associated with rectangular periodic crystals.    
 
\subsubsection*{Acknowledgements}
A.N.N. acknowledges support from  CNRS and from ONR.


\begin{thebibliography}{10}

\bibitem{Strohbook}
J.~J. Wu, T.~C.~T. Ting, and D.~M. Barnett, editors.
\newblock {\em Modern Theory of Anisotropic Elasticity and Applications}.
\newblock SIAM, Philadelphia, 1991.

\bibitem{Ting96}
T.~C.~T. Ting.
\newblock {\em Anisotropic Elasticity: Theory and Applications}.
\newblock Oxford University Press, 1996.

\bibitem{Wu98}
K.-C. Wu.
\newblock Generalization of the {S}troh formalism to 3-dimensional anisotropic
  elasticity.
\newblock {\em J. Elasticity}, 51:213--225, 1998.

\bibitem{Shuvalov00a}
A.~L. Shuvalov.
\newblock On the theory of wave propagation in anisotropic plates.
\newblock {\em Proc. R. Soc. A}, 456:2197--2222, 2000.

\bibitem{Shuvalov04a}
A.~L. Shuvalov, O.~Poncelet, and M.~Deschamps.
\newblock General formalism for plane guided waves in transversely
  inhomogeneous anisotropic plates.
\newblock {\em Wave Motion}, 40(4):413--426, 2004.

\bibitem{Shuvalov03}
A.~L. Shuvalov.
\newblock A sextic formalism for three-dimensional elastodynamics of
  cylindrically anisotropic radially inhomogeneous materials.
\newblock {\em Proc. R. Soc. A}, 459(2035):1611--1639, 2003.

\bibitem{Norris10}
A.~N. Norris and A.~L. Shuvalov.
\newblock Wave impedance matrices for cylindrically anisotropic radially
  inhomogeneous elastic materials.
\newblock {\em Q. J. Mech. Appl. Math.}, 63:1--35, 2010.

\bibitem{Hu54}
H.~C. Hu.
\newblock {On the general theory of elasticity for a spherically isotropic
  medium}.
\newblock {\em Acta Sci. Sin.}, 3:247--260, 1954.

\bibitem{Shulga88}
N.~A. Shul'ga, A.~Y. Grigorenko, and T.~L. Efimova.
\newblock {Free non-axisymmetric oscillations of a thick-walled,
  nonhomogeneous, transversally isotropic, hollow sphere}.
\newblock {\em Int. Appl. Mech.}, 24(5):439--444, 1988.

\bibitem{Chen01}
W.~Chen and H.~J. Ding.
\newblock {Free vibration of multi-layered spherically isotropic hollow
  spheres}.
\newblock {\em Int. J. Mech. Sci.}, 43(3):667--680, 2001.

\bibitem{Hash09}
S.~Hasheminejad and M.~Maleki.
\newblock {Acoustic wave interaction with a laminated transversely isotropic
  spherical shell with imperfect bonding}.
\newblock {\em Arch. Appl. Mech. (Ingenieur Archiv)}, 79(2):97--112, 2009.

\bibitem{Scandrett02}
C.~Scandrett.
\newblock {Scattering and active acoustic control from a submerged spherical
  shell}.
\newblock {\em J. Acoust. Soc. Am.}, 111:893--907, 2002.

\bibitem{Ting98}
T.~C.~T. Ting.
\newblock The remarkable nature of radially symmetric deformation of
  spherically uniform linear anisotropic elastic solids.
\newblock {\em J. Elasticity}, 53:47--64, 1998.

\bibitem{Lekhnitskii}
S.~G. Lekhnitskii.
\newblock {\em Theory of Elasticity of an Anisotropic Elastic Body}.
\newblock Holden-Day, San Francisco, 1963.

\bibitem{Nair}
S.~Nair.
\newblock {\em Introduction to Continuum Mechanics}.
\newblock Cambridge University Press, New York, 2009.

\bibitem{Lothe76}
J.~Lothe and D.~M. Barnett.
\newblock On the existence of surface-wave solutions for anisotropic elastic
  half-spaces with free surface.
\newblock {\em J. Appl. Phys.}, 47(2):428--433, 1976.

\bibitem{Martin06}
P.~A. Martin.
\newblock {\em Multiple Scattering: Interaction of Time-harmonic Waves with {N}
  Obstacles}.
\newblock Cambridge University Press, New York, 2006.

\bibitem{MorseII}
P.~M. Morse and H.~Feshbach.
\newblock {\em Methods of {T}heoretical {P}hysics, Vol. II}.
\newblock McGraw-Hill, New York, 1953.

\bibitem{fed}
F.~I. Fedorov.
\newblock {\em Theory of Elastic Waves in Crystals}.
\newblock Plenum Press, New York, 1968.

\bibitem{Dassios95}
G.~Dassios and Z.~Rigou.
\newblock Elastic {H}erglotz functions.
\newblock {\em SIAM J. Appl. Math.}, 55(5):1345--1361, 1995.

\bibitem{Pease}
M.~C. Pease.
\newblock {\em Methods of Matrix Algebra}.
\newblock Academic Press, New York, 1965.

\bibitem{Wasow}
W.~Wasow.
\newblock {\em Asymptotic Expansions for Ordinary Differential Equations}.
\newblock Interscience, New York, 1965.

\bibitem{Shuvalov03a}
A.~L. Shuvalov.
\newblock The {F}robenius power series solution for cylindrically anisotropic
  radially inhomogeneous elastic materials.
\newblock {\em Q. J. Mech. Appl. Math.}, 56(3):327--345, 2003.

\bibitem{Antman87}
S.~S. Antman and P.~V. Negron-Marrero.
\newblock The remarkable nature of radially symmetric equilibrium states of
  aeolotropic nonlinearly elastic bodies.
\newblock {\em J. Elasticity}, 18:131?-164, 1987.

\bibitem{Antman01}
S.~S. Antman and T.~C.~T. Ting.
\newblock Anisotropy consistent with spherical symmetry in continuum mechanics.
\newblock {\em J. Elasticity}, 62(1):85--93, 2001.

\bibitem{Jolliffe}
A.~E. Jolliffe.
\newblock A form for $\frac{d}{dn} {P}_n(\mu )$, where ${P}_n(\mu )$ is the
  {L}egendre polynomial of degree $n$.
\newblock {\em Mess. Math.}, 49:125--127, 1919.

\bibitem{Torrent09}
D.~Torrent and J.~S\'{a}nchez-Dehesa.
\newblock Radial wave crystals: Radially periodic structures from anisotropic
  metamaterials for engineering acoustic or electromagnetic waves.
\newblock {\em Phys. Rev. Lett.}, 103(6):064301+, 2009.

\end{thebibliography}


\label{lastpage}
\end{document}